\documentstyle[epsf,12pt]{article}
\begin{document}
\pagestyle{empty}
\rightline{\vbox{
\halign{&#\hfil\cr
&OCIP/C-95-7\cr
& May 1995\cr}}}
\bigskip
{\large\bf
\centerline{Fragmentation production of heavy quark states}}
\bigskip
\normalsize
\centerline{Michael A. Doncheski}
\centerline{\sl Department of Physics,
                Carleton University, Ottawa, Ontario K1S 5B6, Canada}
\bigskip
\begin{abstract}
It has recently been noticed that heavy quark ($Q$) and gluon ($g$)
fragmentation to heavy quarkonium states ($Q\bar{Q}$) can be calculated in
perturbative QCD.  This technique allows for the inclusion of a subset of
higher order corrections to quarkonium production which dominates at large
transverse momentum.  A comparison will be presented between the new
calculation of quarkonium production to the existing data from the Tevatron.
In addition some new results on predictions for double heavy quark baryons at
various colliders will be presented.
\end{abstract}
\pagestyle{plain}\setcounter{page}{1}

Our understanding of the production and decays of quarkonium states is not
perfect.  It was supposed that $J/\psi$ production would be dominated by two
sources, the charmonium production model (CPM)\cite{gms} and $B$ meson decay
(BPM)\cite{nde}.  The CDF collaboration measured the $J/\psi$ production cross
section using their 1988-89 data\cite{cdf1}, and found rather poor agreement
with the sum of the two dominant production mechanisms.  They were able to fit
the data to a sum of these two mechanisms if the normalizations were allowed to
float.  This allowed an extraction of the fraction of $J/\psi$ mesons from $B$
meson decay, and that, in turn, allowed for the inclusion of inclusive $J/\psi$
in the measurement of the $b$-quark $p_{_T}$ distribution.  The $J/\psi$ data
points were significantly higher than the theoretical predictions\cite{vaia1},
though the UA1 collaboration found good agreement between theory and
experiment, using similar techniques\cite{huth}.  This lead to many
investigations of the gluon distribution in the proton and the of $b$-quark
production\cite{quest}.

The questions about the source of the disagreement between theory and
experiment were settled when the CDF detector was upgraded to include a silicon
vertex detector (SVX).  With the SVX, it was possible to separate the prompt
$J/\psi$ mesons ({\it i.e.}, those originating at the production vertex) from
those produced in $B$ meson decay ({\it i.e.}, those with a displaced vertex).
Using their 1992-93 data, the CDF collaboration was able to state that there
were far too many prompt $J/\psi$'s, and the number produced in $B$ decay was
in good agreement with theoretical predictions\cite{cdf2}.  Thus the problem
shifted from our understanding of the gluon distribution and $b$-quark
production to that of $J/\psi$ production.

Due to the relatively large mass of the $c$-quark, $\alpha_s$ is small and
heavy quarkonium production and decays can be calculated using perturbative QCD
(pQCD), and these calculations were, of course, performed.  $S$-wave state
production
and decay ($J/\psi$, $\eta_c$, $\ldots$) are well behaved.  However, a
long-standing problem in quarkonium physics that of the hadronic decay (and
similarly in the hadronic production) of $P$-wave states.  It was noticed that
the matrix element of $P$-wave quarkonium states to 3 gluons has a soft
singularity (again, even though the $S$-wave production and 3 gluon decays are
well behaved)\cite{barbieri}.  Reasonable results are obtained if one uses the
binding energy, confinement radius or the radius of the bound state to cut-off
the divergences, but these are completely unjustified and non-rigorous
procedures.

The latter problem was addressed in Ref.~\cite{bbl1} (which was later expanded
into a rigorous theory called Non-Relativistic QCD, or NRQCD\cite{bbl2}).  A
physical hadronic state is made up of a superposition of an infinite number of
Fock states:
\begin{equation}
\mid M \rangle = \psi_{Q\bar{Q}} \mid Q\bar{Q} \rangle + \psi_{Q\bar{Q}g}
\mid Q\bar{Q}g \rangle + \cdots
\end{equation}
Generally, the states with one or more gluons don't contribute significantly,
as the radiation of gluons by heavy quarks $Q$ is suppressed by order $v^2$.
However, in the case of $P$-wave decays, the angular momentum barrier
suppresses the $P$-wave annihilation by order $v^2$, relative to the $S$-wave
annihilation.  As the $Q\bar{Q}$ in the $\mid Q\bar{Q}g \rangle$ state can be
in the $S$-wave, both the $P$-wave color singlet annihilation and the $S$-wave
color octet annihilation are the same order in NRQCD.  Thus the decay of the
$P$-wave physical states is given by:
\begin{equation}
\Gamma(n[^3P_J]\rightarrow hadrons) =
H_1(n) \hat{\Gamma}_1(Q\bar{Q}[^3P_J] \rightarrow partons) +
H_8(n) \hat{\Gamma}_8(Q\bar{Q}[^3S_1] \rightarrow partons)
\end{equation}
where the subscripts 1 and 8 refer to the color singlet and octet,
respectively.  The $H_1$ and $H_8$ are non-perturbative parameters.  The soft
divergence found in earlier works was due to the missing $H_8$ parameter.

\ \vspace{-0.5cm}\\
\begin{minipage}[t]{7.8cm} {
\begin{center}
\hspace{-1.7cm}
\mbox{
\epsfysize=7.0cm
\epsffile[0 -125 500 500]{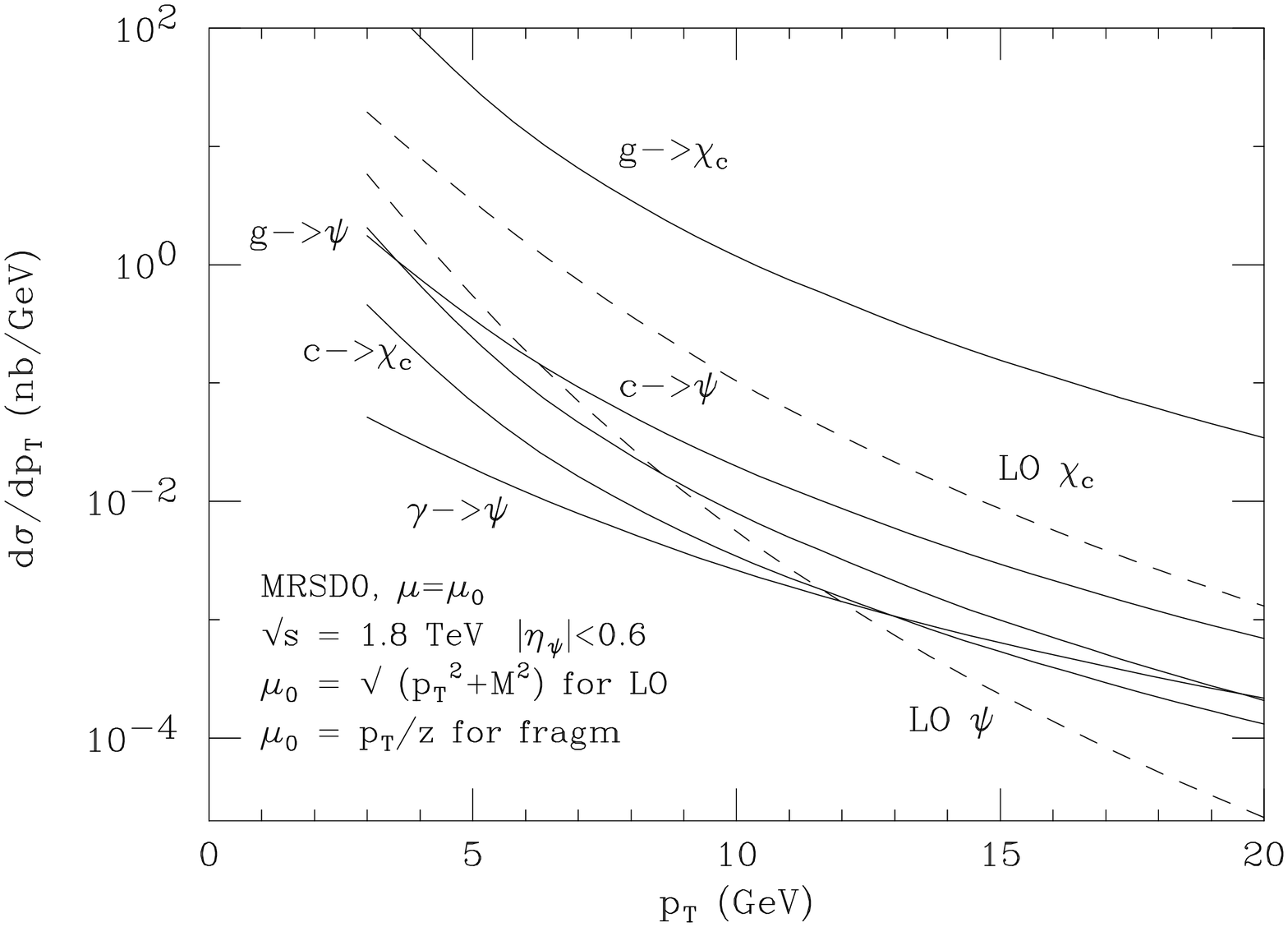}
}
\end{center}
\vspace*{-2.0cm}
\noindent
{\small\bf Fig.~1. }{\small\it
Contributions of the various mechanisms to prompt $J/\psi$ production at the
Tevatron.  The $\chi_{_J}$ contributions include the branching ratio for
radiative decay to $J/\psi$.
}
}\end{minipage}
\hspace{0.5cm}
\begin{minipage}[t]{7.8cm} {
\begin{center}
\hspace{-1.7cm}
\mbox{
\epsfysize=7.0cm
\epsffile[0 -125 500 500]{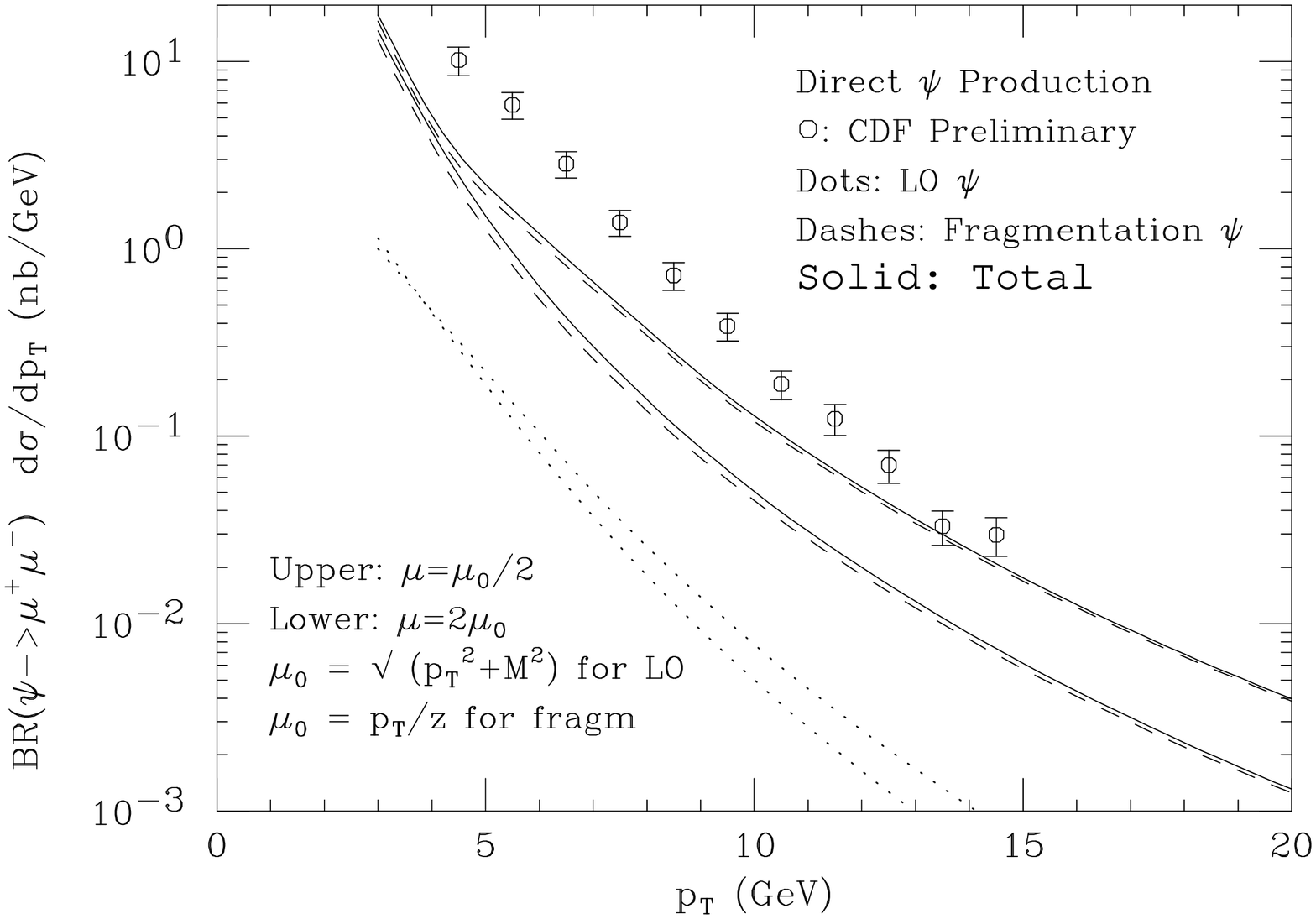}
}
\end{center}
\vspace*{-2.0cm}
\noindent
{\small\bf Fig.~2. }{\small\it
The total prompt $J/\psi$ production rate at the Tevatron.  The direct
(dotted), fragmentation (dashed) and total (solid) contributions are shown
separately; also shown are the CDF preliminary data points.
}
}\end{minipage}
\vspace*{0.5cm}

The solution to the $J/\psi$ production at CDF problem is due to the existence
of previously uncalculated fragmentation contributions.  These contributions
are higher order in $\alpha_s$, and so were thought to be negligible compared
to the direct production mechanisms included in Ref.~\cite{gms}.  However, the
direct production mechanisms required one to calculate $Q\bar{Q}$ production
where the quark-anti-quark pair have both small relative momentum and the
correct quantum numbers to form the relevant bound states.  These restrictions
modify the $p_{_T}$ distributions of the produced states.  For example,
$g+g \rightarrow g+g$ falls off as $1/p_{_T}^4$, while
$g+g \rightarrow J/\psi+g$ falls off as $1/p_{_T}^8$, at large $p_{_T}$.  The
fragmentation of a gluon into a quarkonium state will not modify the $p_{_T}$
distribution significantly.  Including the proper factors of $\alpha_s$, the
large $p_{_T}$ limit of the fragmentation production of $J/\psi$ is of order
$\alpha_s^5/p_{_T}^4$ while direct production goes as $\alpha_s^3/p_{_T}^8$;
similarly, the large $p_{_T}$ limit of $\chi_{_J}$ (the $P$-wave, spin 1
states) production via fragmentation is of order $\alpha_s^4/p_{_T}^4$ while
the direct production goes as $\alpha_s^3/p_{_T}^6$.  At large enough $p_{_T}$,
the slower fall-off with $p_{_T}$ of the fragmentation contributions can make
up for the extra powers of $\alpha_s$.  The correct calculation of the
fragmentation production of $P$-wave states requires the use of NRQCD, and the
color octet contribution is important.  Many fragmentation functions have
involving heavy quark states have been calculated, and those used in for the
following analyses are: gluon fragmentation to quarkonium states\cite{gfrag};
heavy quark fragmentation to quarkonium states\cite{qfrag}; photon
fragmentation to $J/\psi$\cite{gamfrag}; and heavy quark fragmentation to heavy
quark-heavy quark diquarks\cite{falk}.

\ \vspace{-0.5cm}\\
\begin{minipage}[t]{7.8cm} {
\begin{center}
\hspace{-1.7cm}
\mbox{
\epsfysize=7.0cm
\epsffile[0 -125 500 500]{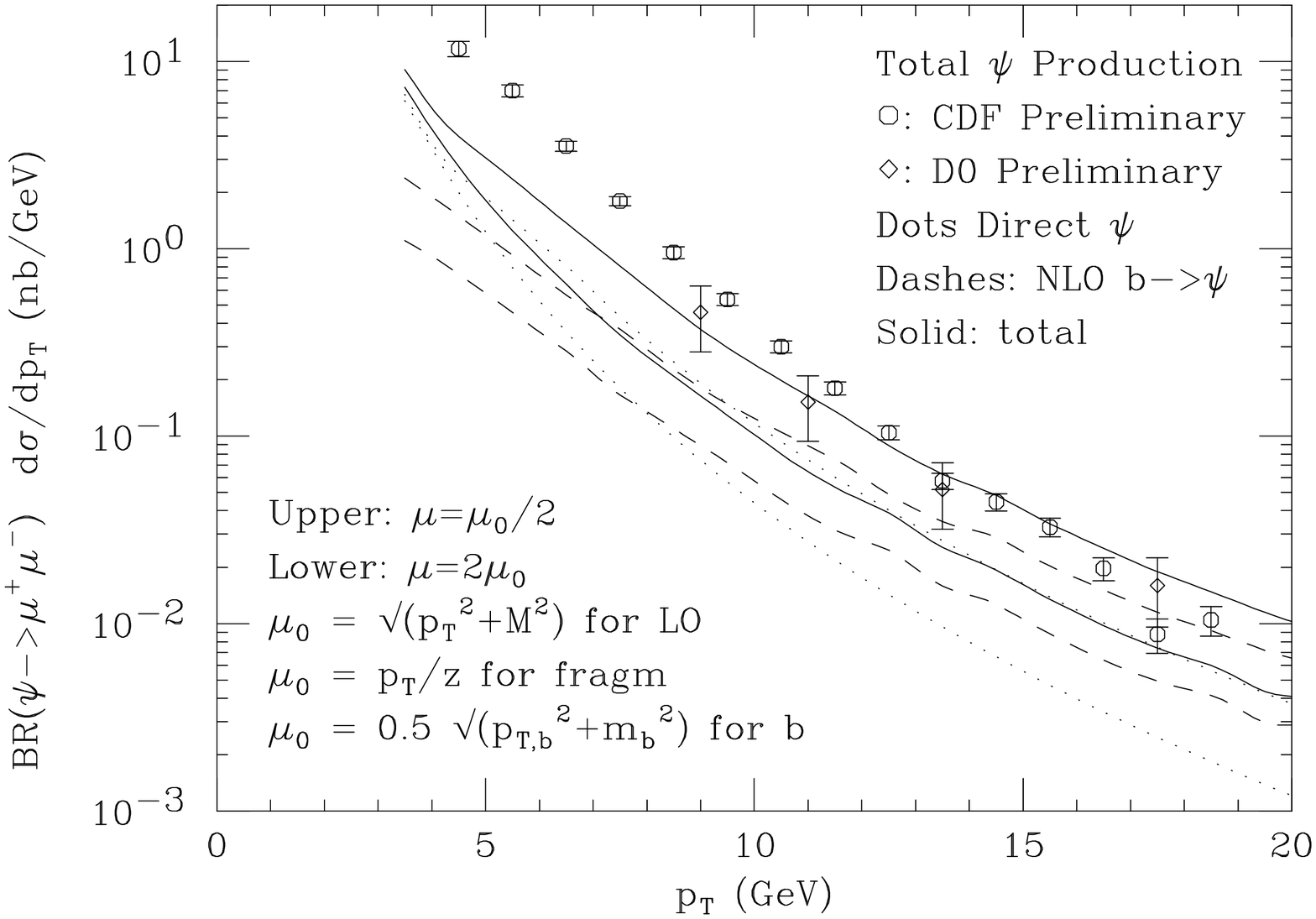}
}
\end{center}
\vspace*{-2.0cm}
\noindent
{\small\bf Fig.~3. }{\small\it
The total $J/\psi$ rate at the Tevatron.  The prompt (dotted), $B$ meson decay
(dashed) and total (solid) contributions are shown; also shown are the CDF and
D0 preliminary data points.
}
}\end{minipage}
\hspace{0.5cm}
\begin{minipage}[t]{7.8cm} {
\begin{center}
\hspace{-1.7cm}
\mbox{
\epsfysize=7.0cm
\epsffile[0 -125 500 500]{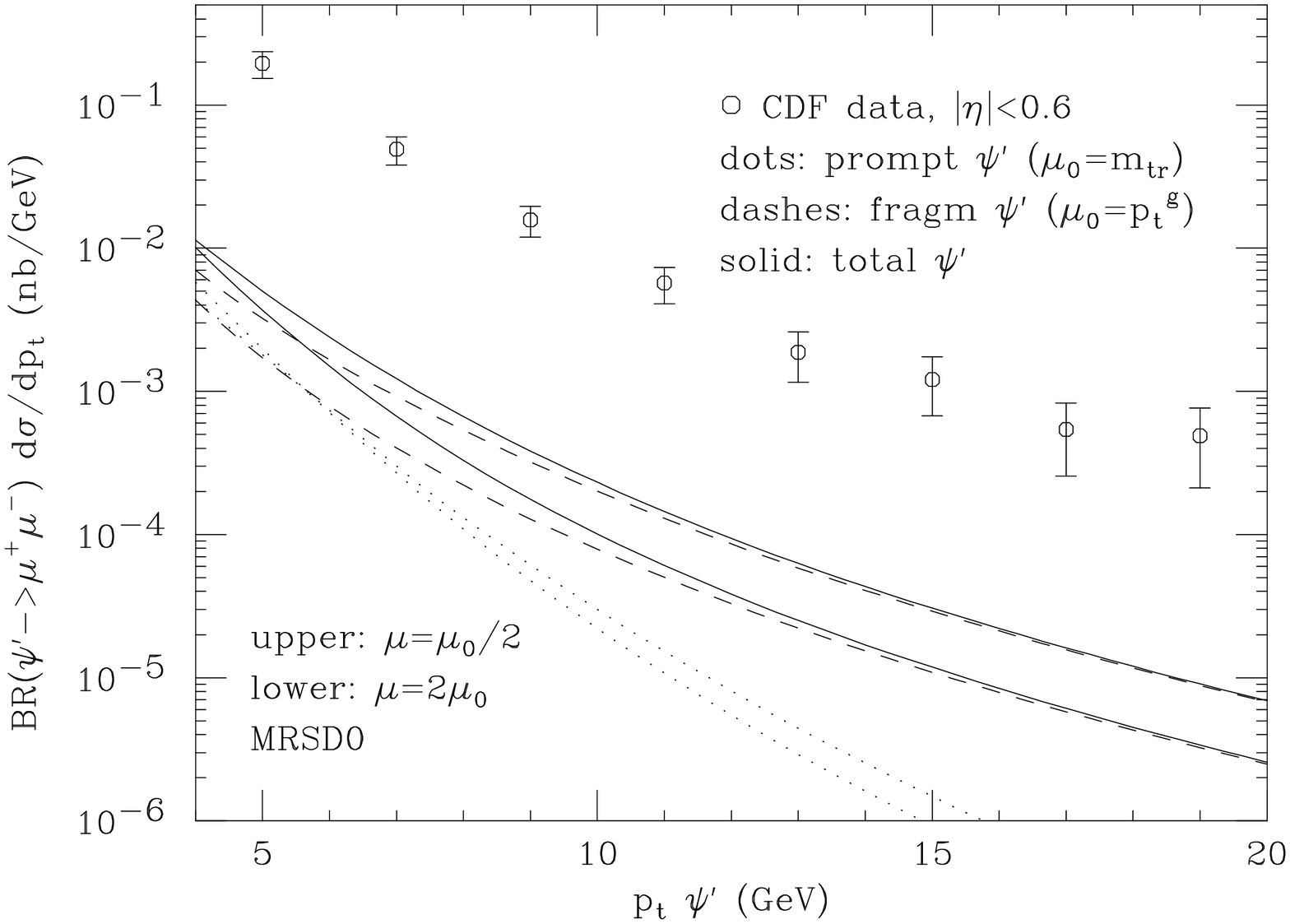}
}
\end{center}
\vspace*{-2.0cm}
\noindent
{\small\bf Fig.~4. }{\small\it
The total $\psi'$ rate at the Tevatron.  The direct (dotted), fragmentation
(dashed) and total (solid) contributions are shown; also shown are the CDF
preliminary data points.
}
}\end{minipage}
\vspace*{0.5cm}

\ \vspace*{-1.5cm}\\
\begin{minipage}[t]{7.8cm} {
\begin{center}
\hspace{-1.7cm}
\mbox{
\epsfysize=8.0cm
\epsffile[0 -125 500 500]{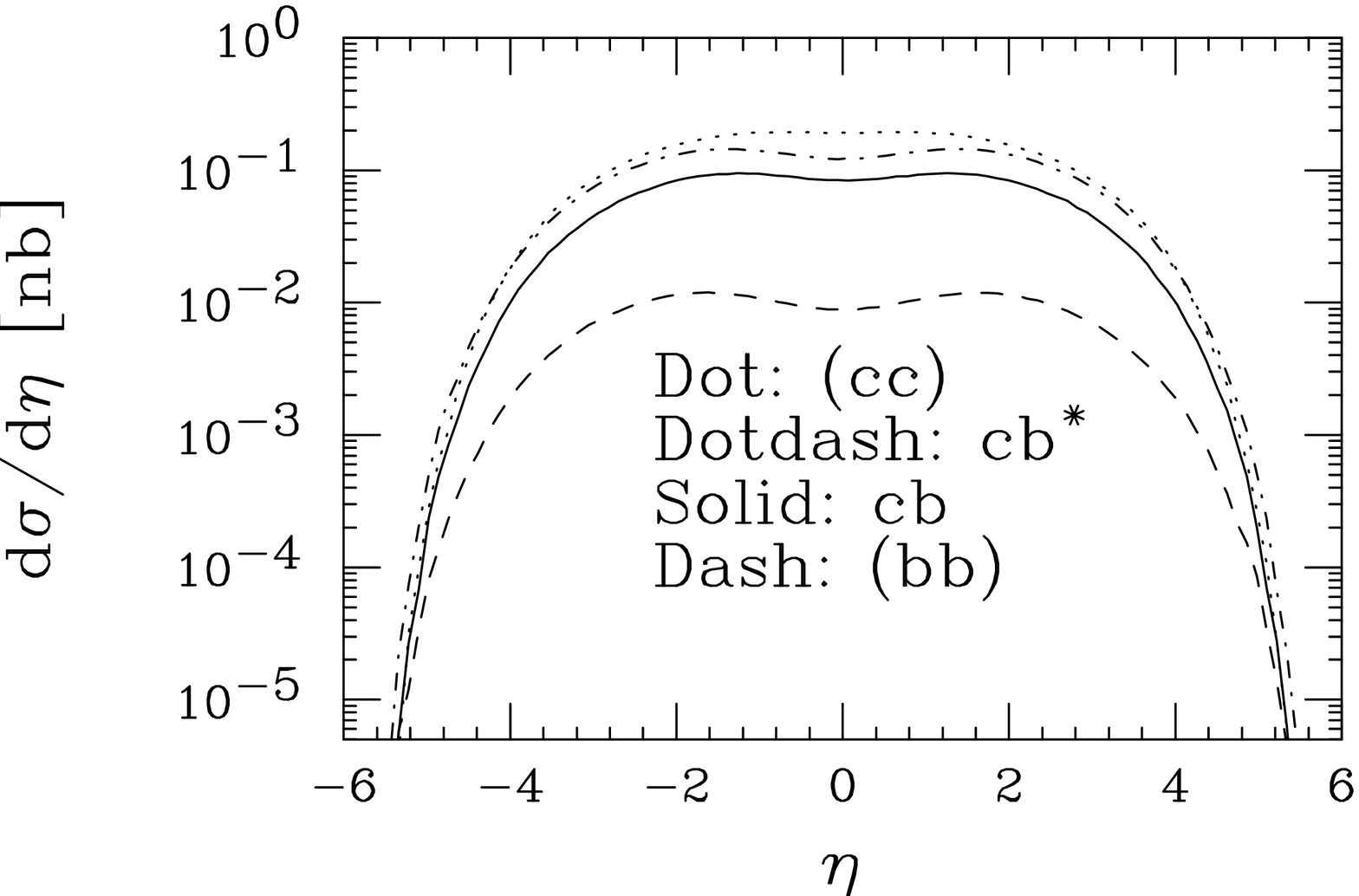}
}
\end{center}
\vspace*{-2.0cm}
\noindent
{\small\bf Fig.~5. }{\small\it
Pseudorapidity distributions for the various contributions to double heavy
quark baryon production at the Tevatron.
}
}\end{minipage}
\hspace{0.5cm}
\begin{minipage}[t]{7.8cm} {
\begin{center}
\hspace{-1.7cm}
\mbox{
\epsfysize=8.0cm
\epsffile[0 -125 500 500]{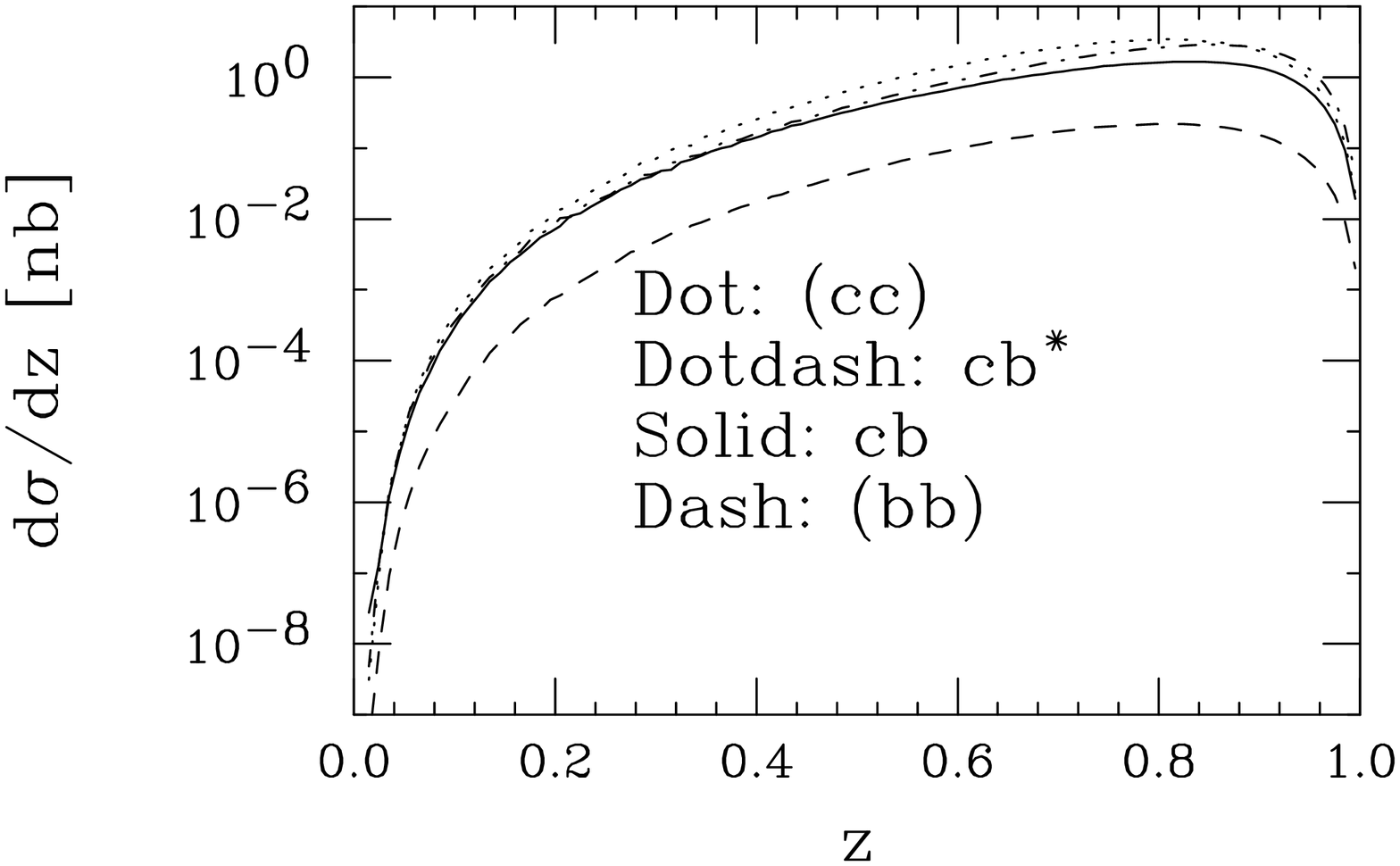}
}
\end{center}
\vspace*{-2.0cm}
\noindent
{\small\bf Fig.~6. }{\small\it
Momentum fraction distributions for the various contributions to double heavy
quark baryon production at the Tevatron.
}
}\end{minipage}
\vspace*{0.5cm}

Now the problem of $J/\psi$ production at the Tevatron may be fully addressed.
Three groups presented compatible results at very nearly the same
time\cite{us1,cacciari,sridhar1}.  Results from Ref.~\cite{us1} will be
presented below.  All $2\rightarrow2$ subprocesses were included;
MRSD0\cite{mrs} parton distribution functions were used, and $\mu = p_{_T}$ (of
the fragmenting parton) for the fragmentation contributions and $\mu = M_{_T}$
(of the $J/\psi$) for the direct contributions unless otherwise noted (here
$\mu$ is the fragmentation, factorization and renormalization scale, all chosen
to be equal); $\mid \eta \mid < 0.6$ was used to simulate the detector
acceptance.  In Fig.~1, the various contributions to prompt $J/\psi$ production
are shown.  In Fig.~2, the total prompt $J/\psi$ production rate is shown; the
scale $\mu$ is varied to show the theoretical uncertainties due to the scale
choice.  The direct contribution is taken from Ref.~\cite{mlm}.  The
theoretical results are compared to the CDF data\cite{cdf2}.  In Fig.~3, the
total $J/\psi$ rate is shown, including both prompt (direct and fragmentation)
and $B$ decay mechanisms; theoretical results are compared to the
CDF\cite{cdf2} and D0\cite{d0} data.  In Fig.~4, the results of $\psi'$ are
shown, compared to the CDF\cite{cdf3} data.  The total estimated theoretical
error (including scale dependence, QCD and relativistic corrections) is of
order a factor of 2.  The $J/\psi$ production rate is now in good agreement
with data; the $\psi'$ discrepancy will be discussed later.

\ \vspace*{-1.5cm}\\
\begin{minipage}[t]{7.8cm} {
\begin{center}
\hspace{-1.7cm}
\mbox{
\epsfysize=8.0cm
\epsffile[0 -125 500 500]{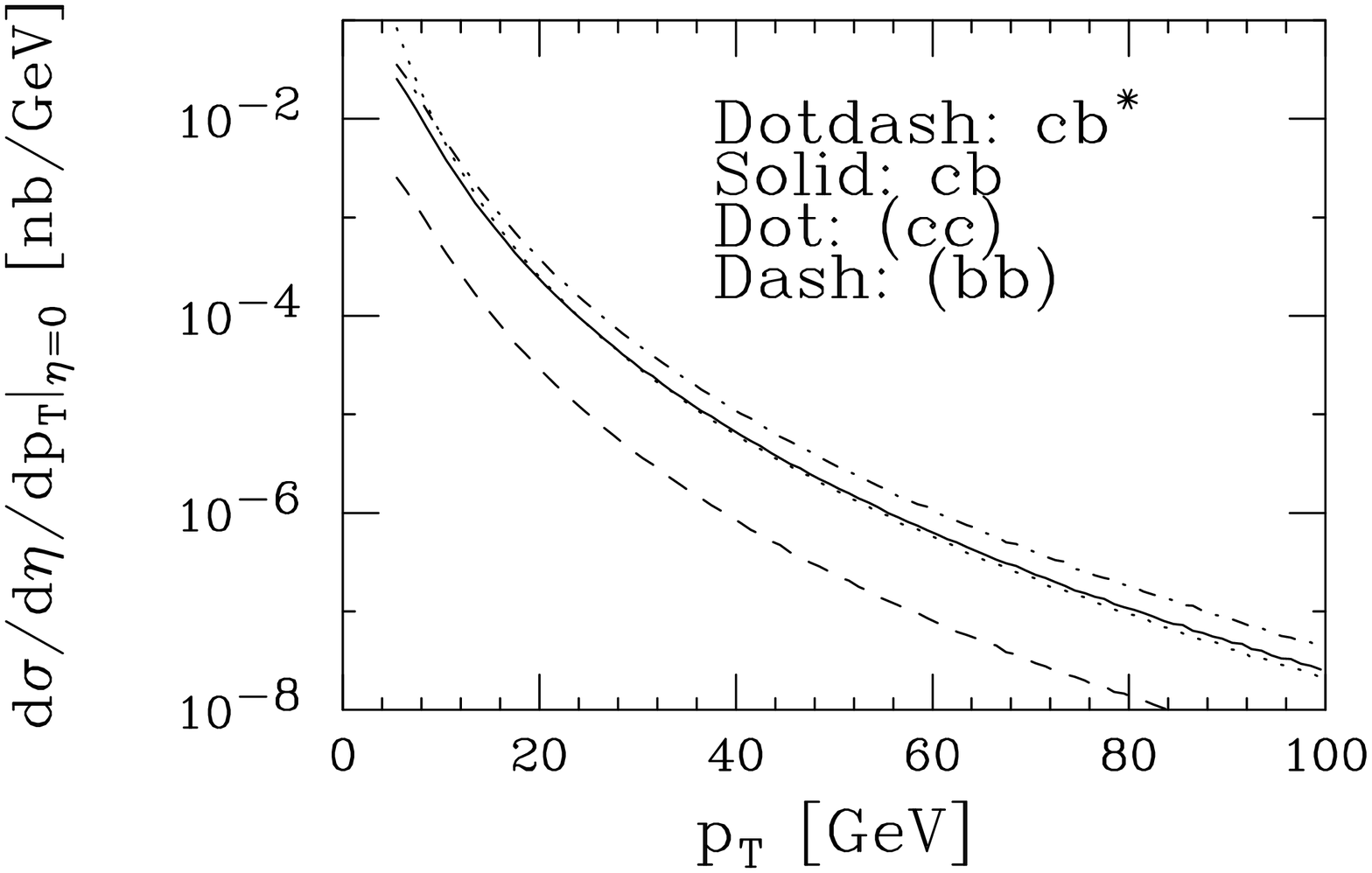}
}
\end{center}
\vspace*{-2.0cm}
\noindent
{\small\bf Fig.~7. }{\small\it
$p_{_T}$ distributions (at small rapidity) for the various contributions to
double heavy quark baryon production at the Tevatron.
}
}\end{minipage}
\hspace{0.5cm}
\begin{minipage}[t]{7.8cm} {
\begin{center}
\hspace{-1.7cm}
\mbox{
\epsfysize=8.0cm
\epsffile[0 -125 500 500]{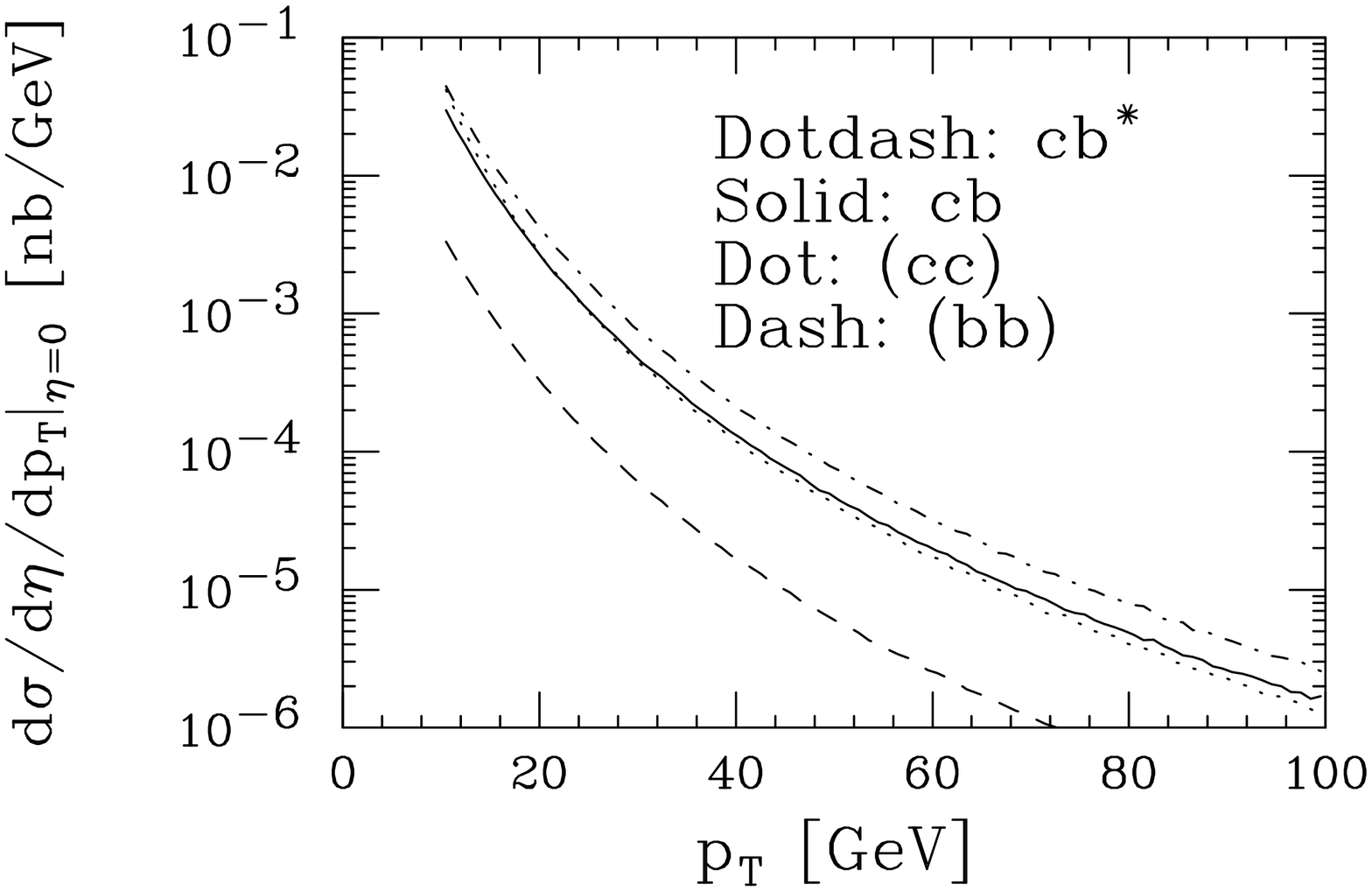}
}
\end{center}
\vspace*{-2.0cm}
\noindent
{\small\bf Fig.~8. }{\small\it
$p_{_T}$ distributions (at small rapidity) for the various contributions to
double heavy quark baryon production at the LHC.
}
}\end{minipage}
\vspace*{0.5cm}

The next set of results to be presented are preliminary, and based upon work in
progress.  Using the fragmentation functions calculated by Falk and
collaborators\cite{falk}, the production rate for double heavy quark diquarks
can be calculated: $c \rightarrow (cc)$; $b \rightarrow (bb)$;
$c,b \rightarrow cb$; and $c,b \rightarrow cb^*$.  The notation is that $(QQ)$
is the spin 1 diquark made up of the same flavor quarks, $QQ'$ is the spin 0
diquark made up of different flavor quarks and $QQ^{\prime *}$ is the spin 1
diquark made up of different flavor quarks.  It is assumed that the diquark
always hadronizes into a double heavy quark baryon.  The fragmentation
probabilities for $c \rightarrow \Sigma_{cc}, \Sigma_{cc}^*$;
$b \rightarrow \Lambda_{bc}$; and $b \rightarrow \Sigma_{bc}, \Sigma~{bc}^*$
are all of order $10^{-5}$, while the remaining probabilities are strongly
suppressed by a factor of $(m_c/m_b)^3$.  The production rates at the NLC
(assumed to be a $60 fb^{-1}/yr$, 500~GeV $e^+e^-$ collider) are small, of
order 30 events/year.  A similar event rate was obtained at HERA.  Event rates
are substantial at both the Tevatron (of order 20k events/year) and the LHC
($10^7$ events/year).  In Fig.~5, the pseudorapidity $\eta$ distributions at
the Tevatron are shown for the various contributions.  In Fig.~6, the momentum
fraction distributions at the Tevatron are shown, demonstrating the strong
peaking at large $z$ ($z$ is defined to be the ratio of the momentum of the
double heavy quark hadron to that of the fragmenting heavy quark).  Similar
results on the $\eta$ and $z$ distributions are obtained for the LHC, with the
primary difference being a smaller relative contribution of the process
$c \rightarrow (cc)$ at the LHC.  In Figs.~7 and 8, $d\sigma /dp_{_T} /d\eta$
at $\eta=0$ is shown at the Tevatron and the LHC, respectively.  Reconstruction
of these states will pose a challenge to the experimentalists.  Work on this
project is continuing, and final results will soon be ready.

\ \vspace{0cm}\\
\begin{minipage}[t]{7.8cm} {
\begin{center}
\hspace{-1.7cm}
\mbox{
\epsfysize=7.0cm
\epsffile[0 0 500 500]{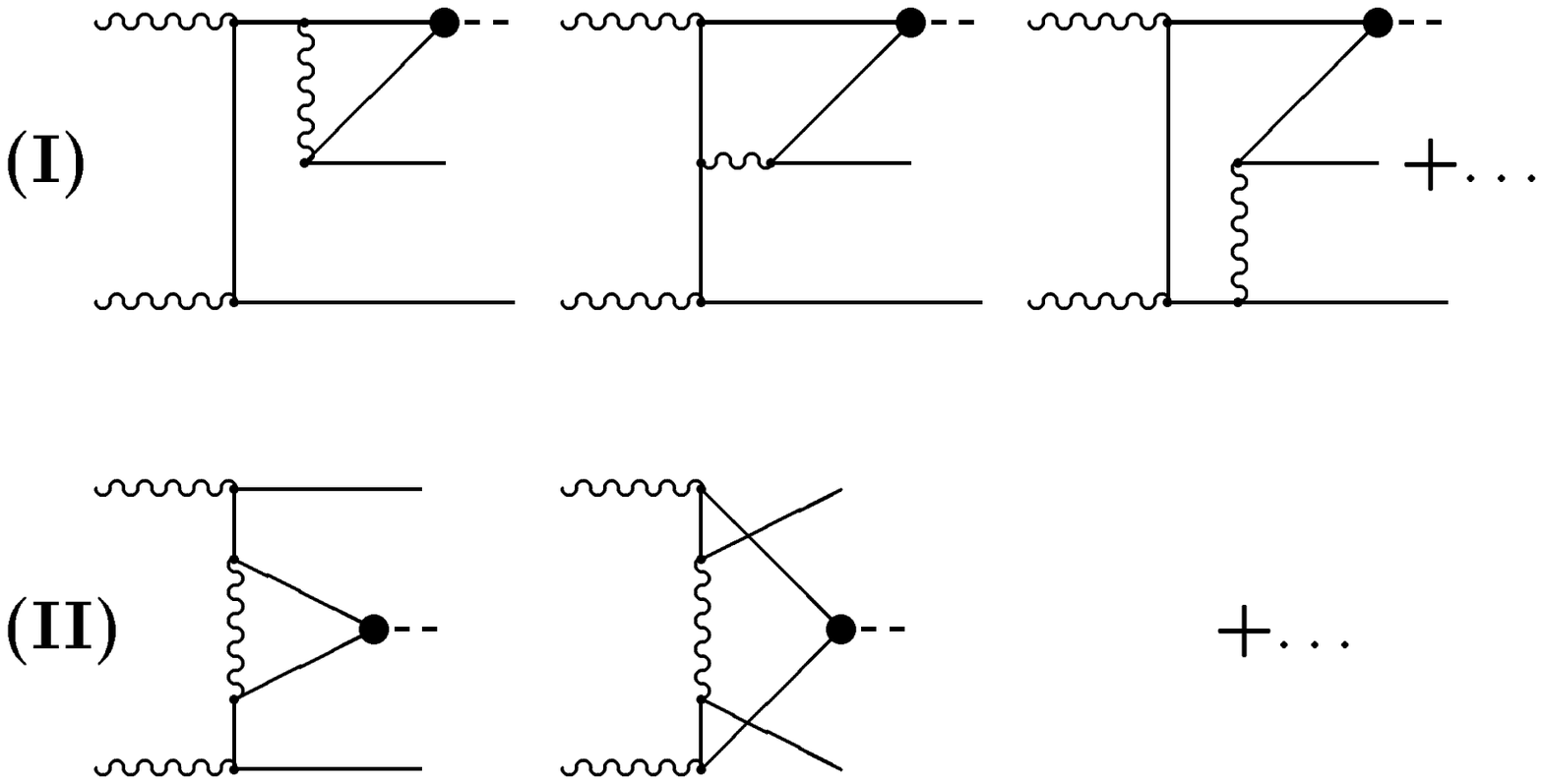}
}
\end{center}
\vspace*{-0.5cm}
\noindent
{\small\bf Fig.~9. }{\small\it
Different topologies of the lowest-order Feynman diagrams contributing
to $\gamma \gamma \rightarrow B_c b \bar{c}$.
}
}\end{minipage}
\hspace{0.5cm}
\begin{minipage}[t]{7.8cm} {
\begin{center}
\hspace{-1.7cm}
\mbox{
\epsfysize=7.0cm
\epsffile[0 0 500 500]{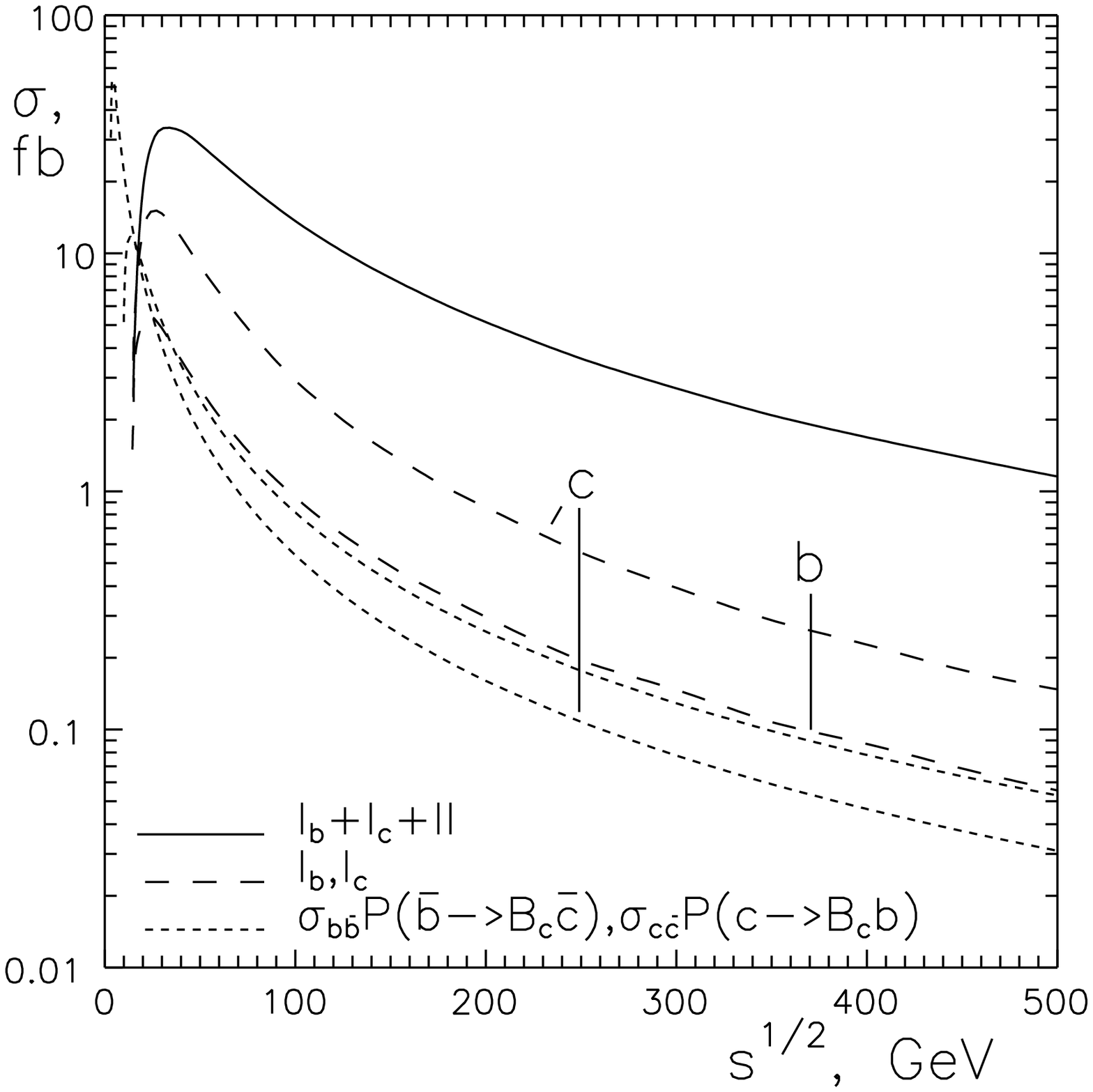}
}
\end{center}
\vspace*{-0.5cm}
\noindent
{\small\bf Fig.~10. }{\small\it
Integrated cross sections for $\gamma\gamma \rightarrow B_c b \bar{c}$
versus the c.m. energy. The different calculations are explained in the text.
The production mechanisms are as classified in Fig.~9.
}
}\end{minipage}
\vspace*{0.5cm}

The authors of Ref.~\cite{kol1} calculated the production of $B_c$ mesons in
photon-photon collisions, using the Feynman diagrams shown in Fig.~9 (the set
(I) can be subdivided into (I$_b$) and (I$_c$) depending on which flavor quark
is connected to the photons).  They find that the recombination diagrams (set
(II)) contribute strongly, and that in this case, the heavy quark state
production can not be well described by a fragmentation process.  Fig.~10 shows
a comparison of the cross section for the full calculation, and for sets
(I$_b$) and (I$_c$) separately, as a function of center of mass energy.
Furthermore, the authors of Ref.~\cite{kol1} compare the results of set (I) to
the relevant heavy quark fragmentation result; the $c$ fragmentation
contribution does not agree well with set (I$_c$) while the $b$ fragmentation
contribution is in relatively good agreement with set (I$_b$).  The inclusion
of the additional diagrams for the gluon-gluon fusion into $B_c$, as well as
the convolution of the parton level subprocess with parton distributions may
change the conclusions at a hadron collider.  It should be noted that the
conclusions of Ref.~\cite{kol1} apply to heavy quark fragmentation
contributions only, and so may affect the production rate of double heavy quark
baryons; the results on $J/\psi$ production {\it via} fragmentation are
dominated by {\em gluon} fragmentation, and the argument of Ref.~\cite{kol1} do
not apply.  After this presentation, another paper discussing the production of
$B_c$ mesons in hadron colliders appeared\cite{kol2}.  The parton level cross
sections ($g+g \rightarrow B_c+X$) are still found to be significantly
different, depending on whether the full calculation or just the fragmentation
approximation is used.  After convoluting the parton level cross sections with
the gluon distributions in the proton, the fragmentation approximation is found
to be in good agreement with the full calculation for $p_{_T} > 10$~GeV.  It is
thus likely that the results presented here on double heavy quark baryon
production at the Tevatron and at the LHC will be in quite good agreement with
a full (though much more complicated) calculation.

There remains a large discrepancy between the theoretical predictions for
$\psi'$ production and the CDF experimental data (see Fig.~4).  As can be seen
from Fig.~1, $J/\psi$ production is dominated by $\chi_{_J}$ production
followed by the radiative decay $\chi_{_J} \rightarrow J/\psi + \gamma$.  The
$\psi'$ is more massive than the $\chi_{_J}$ family, and it is expected that
the radial excitation $\chi'_{_J}$ family will be above open charm threshold.
However, the fall-off of the observed $p_{_T}$ distribution of $\psi'$ appears
to be consistent with a fragmentation type production and completely
inconsistent with a conventional, hard ({\it i.e.}, direct) production
mechanism.  Several authors suggest a {\em metastable} $\chi'_{_J}$, produced
dominantly {\it via} gluon fragmentation and decaying with large branching
ratio to $\psi'$ as a plausible solution to this
problem\cite{sridhar2,close,cho1}.  A search strategy for these $\chi'_{_J}$ in
$B$ meson decays at the CLEO II detector has been proposed\cite{tuan}.
Similarly, conventional $c\bar{c}$ states above open charm threshold, but with
quantum numbers forbidding the decay into $D\bar{D}$ have been proposed as a
possible solution\cite{close}, as have hybrid $c\bar{c}g$ states\cite{close}.
These states would also be produced dominantly {\it via} gluon fragmentation,
and decay with large branching ratio to $\psi'$.  Finally, it has been
suggested that color octet $c\bar{c}$ can be produced {\it via} fragmentation,
and these $c\bar{c}$ can evolve nonperturbatively to a $\psi'$ plus
hadrons\cite{fleming}.

After this presentation, Ref.~\cite{cho2} appeared.  It generalizes the
fragmentation mechanism and includes additional terms from the NRQCD
expansion.  Their results are in excellent agreement with data for $J/\psi$,
$\psi'$ and the lowest three $\Upsilon$ states.

The author would like to thank NSERC (Natural Sciences and Engineering Research
Council - Canada) for support, the organizers of MRST at the University of
Rochester for the opportunity to present this talk and the authors of
Ref.~\cite{kol1} for allowing the use of their Figs.~1 and 2 (which became
Figs.~9 and 10 in this report).

\end{document}